# $AB_2X_4$ spinel structures: similarity and difference between the centrosymmetric, *Fd$\bar{3}$m*, and non-centrosymmetric, *F*$4_1$32, space groups

**Alla Arakcheeva.[a] Arnaud Magrez[a] & Gervais Chapuis [b]**

[a] École Polytechnique Fédérale de Lausanne, SB, IPHYS, Crystal Growth Facility, Lausanne 1015, Switzerland

[b] École Polytechnique Fédérale de Lausanne, SB, IPHYS, Cubotron, Lausanne 1015, Switzerland

**Synopsis** Both centrosymmetric, *Fd$\bar{3}$m*, and non-centrosymmetric, *F*$4_1$32, space groups cannot be distinguished in solving, refining and describing $AB_2X_4$ spinel structures in the harmonic approximation. Only the use of the 3rd order ADP approximation can distinguish between them.

**Abstract** Many compounds belonging to the spinel $AB_2X_4$ structure play an important role due to their wide range of practical applications. Most of them are traditionally assigned to the centrosymmetric space group *Fd$\bar{3}$m*. However, the physical properties of some spinels are incompatible with centrosymmetry. This discrepancy is often accounted for by reducing the symmetry to the non-centrosymmetric space group *F$\bar{4}$3m*, allowing thus small atomic displacements from their original position in *Fd$\bar{3}$m*. In this work, we demonstrate that the loss of the inversion symmetry can occur without any atomic displacements, since the centrosymmetric *Fd$\bar{3}$m* and non-centrosymmetric *F*$4_1$32 space groups are equivalent for structure determination and refinement based on X-ray diffraction data. If consistent with experiment, only the use of an anharmonic model of atomic displacements can distinguish these space groups. This study aims to clarify certain misconceptions regarding the structural symmetry and physical properties of spinel type compounds.

**Keywords: $AB_2X_4$ spinel structures; *F*$4_1$32 and *Fd$\bar{3}$m* space groups; Lattice complexes; Centro- and non-centrosymmetric structures.**

## 1. Introduction

The spinel structure type is commonly adopted by many $AB_2X_4$ compounds, including those with important practical applications (Srikala, 2024; Wang *et al.*, 2023; Rafi *et al.*, 2025; Arshad *et al.*, 2024; He *et al.*, 2023; Song *et al.*, 2023; Shan *et al.*, 2023; Xu *et al.*, 2023; Tsurkan et al., 2021; Narang & Pubby, 2021; Szablowski *et al.*, 2025; and many others). Determining the correct space group symmetry is essential for predicting, understanding and tuning their physical properties. Traditionally, the spinel type $AB_2X_4$ compounds are assigned to the centrosymmetric space group *Fd$\bar{3}$m*. However, doubts have arisen for synthetic $MgAl_2O_4$ where inconsistencies between observed physical properties and centrosymmetric symmetry have been reported (Grimes *et al.*, 1983). Such doubts primarily concern the presence or absence of an inversion centre.

A confirmation of a non-centrosymmetric symmetry, $F\bar{4}3m$, has been published for $MgAl_2O_4$ (Grimes *et al.*, 1983), $ZnFe_2O_4$ (Dronova *et al.*, 2022) among others. Furthermore, two reports refer to the non-centrosymmetric space group $F4_132$ for $AB_2X_4$ compounds. One of these concerns $ZnFe_2O_4$ (Dronova *et al.*, 2022), where the observed *h*00 reflection condition ($h \neq 4n$ with integer *n*) is inconsistent with this space group. The other case involves $LiMn_{1.5}Ni_{0.5}O_4$, in which A = Li, $B_2 = (Mn_{1.5}Ni_{0.5})$ and $X_4 = O_4$ in the general formula $AB_2X_4$ (Amin *et al.*, 2020). In that study, the authors proposed a phase transition from $Fd\bar{3}m$ to its maximal "translationengleiche" subgroup $F4_132$, preserving the same translational charateristics. The A, B and O atomic positions remain identical in both space groups when, origin choice 1 (with $\bar{4}3m$ at the origin) is adopted for $Fd\bar{3}m$.

In this report, we demonstrate that the space groups $Fd\bar{3}m$ and $F4_132$ are equivalent for the modelling and refinement of spinel-like $AB_2X_4$ structures based on conventional X-ray diffraction data. In addition, we demonstrate how a non-centrosymmetric model can be applied and refined by carefully selecting the appropriate experimental X-ray wavelength, and incorporating at least third order ADP's in the structure model.

## 2. Comparison of the space groups $Fd\bar{3}m$ and $F4_132$ for the $AB_2X_4$ spinel structures

### 2.1. Structural parameters

The cubic unit cell of the spinel-like $AB_2X_4$ structure (Figure 1) contains Z = 8 formula units: comprising 32 X atoms, 8 A atoms and 16 B atoms. In both space groups $Fd\bar{3}m$ and $F4_132$, all atoms occupy identical Wyckoff positions with the same multiplicity and atomic coordinates: X – 32*e* (*x, x, x* + symmetry equivalents); A – 8*a* (0, 0, 0 + symmetry equivalents); B – 16*d* (5/8, 5/8, 5/8 + symmetry equivalents). The complete sets of atomic coordinates for each crystallographic site in both space groups are provided in Table 1.

Thus, based on structural parameters alone, no distinction can be made between the $Fd\bar{3}m$ and $F4_132$ space groups. In terms of lattice complexes (Fischer & Koch, 2002), the crystal structures are also identical in both cases: A – 8*a* $\bar{4}3m$ $Fd\bar{3}m$ *a* D; B – 16*d* $.\bar{3}m$ $Fd\bar{3}m$ *c T*; X – 32*e* .3*m* $Fd\bar{3}m$ *e* ..2*D*4*xxx*.

### 2.2. Reflection conditions

The general reflection conditions differ slightly between $Fd\bar{3}m$ and $F4_132$. Specifically, the condition $k + l = 4n$ (where *n* is an integer) is characteristic of $Fd\bar{3}m$ but not of $F4_132$ (as indicated in Table 1). However, in $F4_132$, additional reflection conditions associated with the A, B and X atomic sites, also satisfying $k + l = 4n$, effectively suppress this distinction.

Therefore, based solely on the reflection conditions, it is not possible to distinguish between the $Fd\bar{3}m$ and $F4_132$ space groups for the $AB_2X_4$ spinel structure.

#### 2.3. Resonance scattering and generalized Debye Waller factor

In general terms, the classical approach to distinguish between centro- and non-centrosymmetric space groups is to take advantage of the resonant scattering effect of X-ray diffraction. In the harmonic approximation, the structure factor can be expressed in the following form:

$$F_0(\mathbf{h}) = \sum_j f_j^0(|\mathbf{h}|)g_j \exp(i\delta_j(|\mathbf{h}|)) \exp(2\pi i\mathbf{h}\cdot\mathbf{x_j})T_0(\mathbf{h})$$

Where

$$T_0(\mathbf{h}) = \exp\left(-2\pi^2\sum_{j=1}^{3}\sum_{l=1}^{3} h_j\beta^{jl}h_l\right)$$

is the Debye Waller term in the harmonic approximation. Here $\beta^{jl}$ is related to $U^{jl}$ by the equality

$$U^{jl} = \beta^{jl}/2\pi^2 a^j a^l$$

The scattering factor and resonant terms $f'$ and $f''$ for atom $j$ are expressed in the following relations:

$$f_j = f_j^0 + f_j' + if_j'' = f_j^0 g_j \exp(i\delta_j)$$

And

$$g\cos\delta = \frac{f^0 + f'}{f^0}; \qquad g\sin\delta = \frac{f''}{f^0}$$

In the presence of resonant scattering, the non-centrosymmetric space group, the relation $I(\mathbf{h})$ and $I(-\mathbf{h})$ are different whereas in the centrosymmetric case both intensities are identical. We can calculate the four terms $F(\mathbf{h})$ and its conjugate $F^*(\mathbf{h})$, and $F(-\mathbf{h})$ and its conjugate $F^*(-\mathbf{h})$ and obtain

$$F(\mathbf{h}) = \sum_j f_j^0(|\mathbf{h}|)g_j \exp(i\delta_j(|\mathbf{h}|)) \exp(2\pi \mathrm{i}\mathbf{h}\cdot\mathbf{x_j})$$

$$F^*(\mathbf{h}) = \sum_j f_j^0(|\mathbf{h}|)g_j \exp(-i\delta_j(|\mathbf{h}|)) \exp(-2\pi \mathrm{i}\mathbf{h}\cdot\mathbf{x_j})$$

$$F(-\mathbf{h}) = \sum_j f_j^0(|\mathbf{h}|)g_j \exp(i\delta_j(|\mathbf{h}|)) \exp(-2\pi \mathrm{i}\mathbf{h}\cdot\mathbf{x_j})$$

$$F^*(-\mathbf{h}) = \sum_j f_j^0(|\mathbf{h}|)g_j \exp(-i\delta_j(|\mathbf{h}|)) \exp(2\pi \mathrm{i}\mathbf{h}\cdot\mathbf{x_j})$$

The intensities $I(\mathbf{h})$ and $I(-\mathbf{h})$ are proportional to the products $F(\mathbf{h})F^*(\mathbf{h})$ and $F(-\mathbf{h})F^*(-\mathbf{h})$ respectively. We can see that the products $F(\mathbf{h})F^*(\mathbf{h})$ and $F(-\mathbf{h})F^*(-\mathbf{h})$ are different by concentrating on the signs of the two exponential terms containing $\delta_j$ and $\mathbf{h}\cdot\mathbf{x}_j$. In the first product *i.e.* $I(\mathbf{h})$ we have the signs ++ respectively - - whereas in the second product, i.e. $I(-\mathbf{h})$ we have the exponential signs +- respectively -+ which are different from the first product. Consequently, we can conclude that for non-centrosymmetric structures $I(\mathbf{h})$ and $I(-\mathbf{h})$ are different.

One may wonder if this difference in intensities could be exploited in our spinel example? Unfortunately, this is not the case. The reason is that in both centro- and non-centrosymmetric cases, each pair of identical atoms A, B and X belong to identical lattice complexes in both space groups. In other words, the two space groups, centrosymmetric and non-centrosymmetric, cannot be distinguished in the *harmonic* approximation of the diffraction model and this is easily confirmed by simulations.

There is however some possibility offered by diffraction to distinguish between the centro- and non-centrosymmetric space groups. In the presence of anharmonicity, the Debye Waller factor can be generalized by including higher order terms in the power series. If we limit to the third order terms, we obtain the following generalized expression of the Debye-Waller term (Trueblood *et al.*, 1996):

$$T(\mathbf{h}) = T_0(\mathbf{h})[1 + (2\pi i)^3 \gamma^{jkl} h_j h_k h_l / 3!]$$

Here $\gamma^{jkl}$ are the components of a third order tensor and for simplification we assume that the summations over the three pairs of identical indices are implied.

Table 2 shows that the independent ADP parameters are identical for the $Fd\bar{3}m$ and $F4_132$ space groups up to 2nd order tensor approximations for each of A, B, X atom. Using the 3rd order ADP tensor allows them to be distinguished, provided, of course, that an optimized X-ray wavelength has been selected and that the sensitivity of the X-ray diffraction and the quality of the data are sufficient. The reason concerns the difference in the 3rd order independent ADP tensor parameters for atoms X and B: namely, $C_{112} = C_{133} = C_{223} \neq C_{113} \neq 0$ in $F4_132$, while $C_{112} = C_{133} = C_{223} = C_{113} \neq 0$ in $Fd\bar{3}m$ for X; $C_{113} \neq 0$ in $F4_132$, while $C_{113} = 0$ in $Fd\bar{3}m$ for B. Based on the simulation of the $LiMo_2O_4$ structure (see 2.4. Example) using Co and Mo radiation (Table 3), Table 4 shows the influence of the choice of wavelength, since the $f''$ value is decisive for F(**h**). For the B atom (B = Mn in the example), as the heaviest, this value primarily influences F(**h**). Another question arises: is there really an anharmonic shift for the B atom? The answer depends on the compound. Figure 2 shows the dependence of the probability distribution function (p.d.f.) on the possible anharmonic contribution $C_{113}$ of the B = Mn atom in the simulated $LiMn_2O_4$ structure. But it is impossible to predict whether the $C_{113}$ term for B is realistic for a particular $AB_2X_4$ compound without experimental data.

#### 2.4. Example

To simulate the structure of $LiMn_2O_4$ in the centrosymmetric, $Fd\bar{3}m$, and non-centrosymmetric, $F4_132$, space groups, the JANA2006 software package (Petříček et al., 2014) was used for two radiations, Mo and Co, to highlight the influence of the choice of wavelength on the results. The simulations were performed using ADP parameters up to the 3rd order tensor. The same set of atomic coordinates in the same Wyckoff positions was used. The ADP parameters up to the 2nd order tensor were fixed at the same values in both space groups. Other details of the structure simulation are given in Table 3.

Table 4 summarizes the results of the structure simulations. No difference between squared structure factor $|F(\mathbf{h})|^2$ and $|F(-\mathbf{h})|^2$ can be observed in the centrosymmetric space group $Fd\bar{3}m$. However, in non-centrosymmetric $F4_132$ space group, small differences between $|F(\mathbf{h})|^2$ and $|F(-\mathbf{h})|^2$ appear. This difference is higher when Co radiation with anomalous scattering $f''$(Mn) = 3.555 is applied in comparison to Mo radiation with $f''$(Mn) = 0.728. This points to the importance of optimizing the wavelength selection if the result needs to discriminate between centrosymmetric and non-centrosymmetric space groups.

### 3. Discussion

The case of $LiMn_{1.5}Ni_{0.5}O_4$ (Amin *et al.*, 2020) is an interesting one showing the ambiguities resulting from the arbitrary selection of the origins of the specific space groups. This occur frequently while describing series of parent structures in the presence of phase transitions. The selection of the origins of space groups in the International Tables is based on theoretical considerations which are independent of the behaviour of chemical compounds under considerations.

A convenient way to represent sequences of phase transition is often based on the Bärnighausen tree principle. The tree structure follows a sequence of maximal subgroups of the space groups which are of two types, either klassen- or translationengleich. We can clearly illustrate our point with the spinel structure $LiMn_{1.5}Ni_{0.5}O_4$.

The first transformation from $Fd\bar{3}m$ (No 227) to $F4_132$ (No. 210) indicates very different atomic coordinates (Table 5 according to Fig. 3 in the mentioned publication) which at a first glance hints to important structural changes. Once we realise that the higher symmetry structure is described with origin choice 2 (not indicated in the mentioned publication) and that the lower symmetry one is described in an origin setting which is parent to choice 1 in $Fd\bar{3}m$ (Table 6), we find that the two structures are indistinguishable! The spinel structure can be described and presented by four conventional sets of atomic coordinates. Two of them correspond to two choices of origin in $Fd\bar{3}m$: (i) at $\bar{4}3m$ (origin choice 1) and (ii) at $\bar{3}m$ (origin choice 2). For each origin choice, two sets of atomic coordinates can be chosen; they are shifted by (½ ½ ½). Of course, the atomic coordinates are related to each other for these origins. Table 6 and Figure 3 illustrate the identity of these choices.

In the same figure 3 of the mentioned publication, the confusion is also present in the reordering transition from space group $F4_132$ (No 210) to $P4_132$ with a maximal subgroup of index k4 (k for klassengleich). Here again we would expect small shifts in the atomic parameters. In reality the large difference in the atomic coordinates results from the arbitrariness of the position of the origin of $P4_132$ described in the International Tables for Crystallography. Thus, while comparing different structures, it is always allowed and recommended to shift the origin of some space groups in order to better illustrate the close parenthood between the pair of structures under considerations.
The next important point to discuss is the concept of lattice complexes introduced and described by Fischer and Koch (2002) in the International Tables of Crystallography. Identical lattice complexes are rarely (if ever) found in compounds with low symmetry, but they are indeed very important in compounds with high, especially cubic, symmetry, not only in the spinel group compounds but also in the perovskite group compounds. Indeed, the lattice complex of the perovskite-like $ABX_3$ structures with the centrosymmetric $Pm\bar{3}m$ space group (no. 221) is identical to the lattice complex of the non-centrosymmetric space group $P432$ (no. 207): site A – 1*a* $m\bar{3}m$ $Pm\bar{3}m$ *a P*; site B – 1*b* $m\bar{3}m$ $Pm\bar{3}m$ *a P*; site X – *3c* 4/*m m .m c J*. This means that for perovskite-like compounds, the difference between centrosymmetric and non-centrosymmetric space groups is very difficult to determine based on X-ray experiments alone. The case of perovskite-like compounds is identical to that of spinel.
An additional point for discussion concerns the reflection conditions, which are very important in distinguishing between the $Fd\bar{3}m$ and $F4_132$ space groups. The reflection condition $k + l = 4n$ for $0kl$ is valid for both groups (see Table 1) and only for atoms for which harmonic or anharmonic displacements are absent. According to our simulations using Co-radiation and 3$^{rd}$ order ADP parameters for $LiMn_2O_4$ with $F4_132$ space group, three independent reflections, namely 024, 046 and 028, are really present with their intensities 2.0, 7.2 and 7.2 respectively. For comparison, the maximal and minimal reflection intensities of other reflections are 135809.0 (for $hkl$ = 004) and 32.3 (for $hkl$ = 224), respectively. This means that if the ADP 3$^{rd}$ order parameters are meaningful in the experimental data, the difference between the $Fd\bar{3}m$ and $F4_132$ space groups could be observed. However, in practice, this might be difficult.

#### 4. Conclusions

For spinel-like $AB_2X_4$ compounds, the centrosymmetric $Fd\bar{3}m$ space groups and the non-centrosymmetric $F4_132$ space groups are difficult to distinguish based solely on the X-ray diffraction data. We have seen that in the harmonic approximation, X-ray resonance scattering effect cannot be used to distinguish spinel compounds where each atom type belongs to the same lattice complex in both centrosymmetric and non-centrosymmetric space groups. A possible method is to use anharmonic Debye-Waller tensor terms up to at least 3$^{rd}$ order. Consequently, the physical properties of a particular compound are often decisive in determining the presence of inversion symmetry. Similar ambiguity can also arise in perovskite-like $ABX_3$ compounds, which are also of great practical importance. Furthermore, when comparing different structures, it is always allowed and recommended to shift the origin of some space groups in order to better illustrate the close relationship between the pair of structures under consideration.

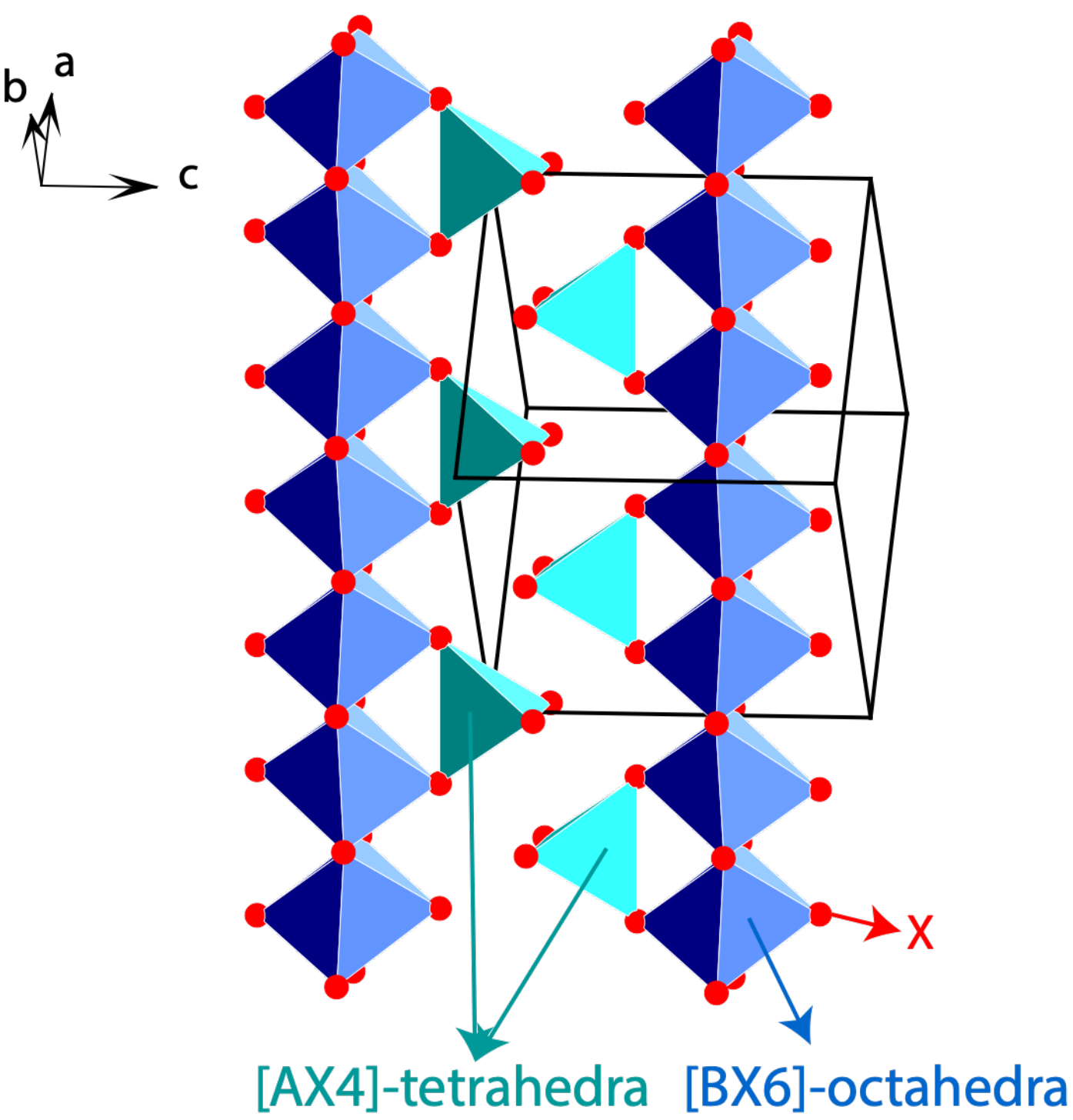

**Figure 1.** Characteristic fragment of $AB_2X_4$ spinel-like structure showing its main structure elements. The cubic unit cell is indicated by black lines.

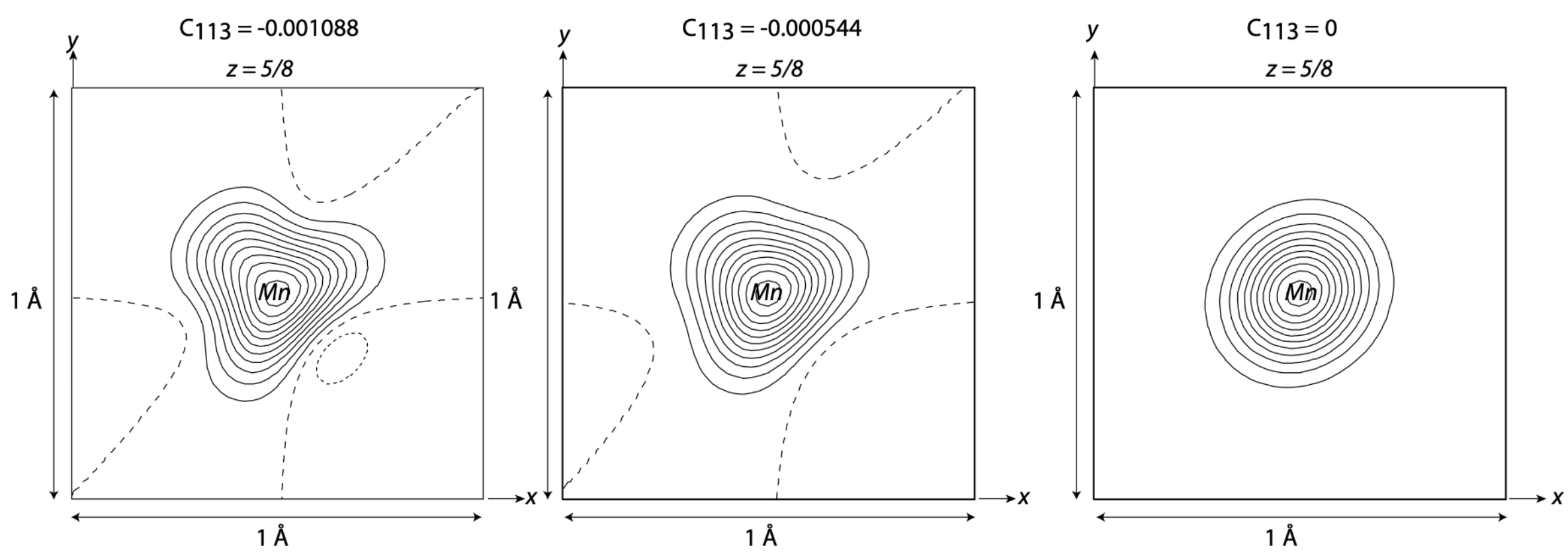

**Figure 2.** Illustration of the probability distribution function (p.d.f.) obtained using the $C_{113}$ term of the 3$^{rd}$ order ADP tensor for Mn in the simulated spinel structure $LiMn_2O_4$ using Co radiation. The same scale is used for all maps.

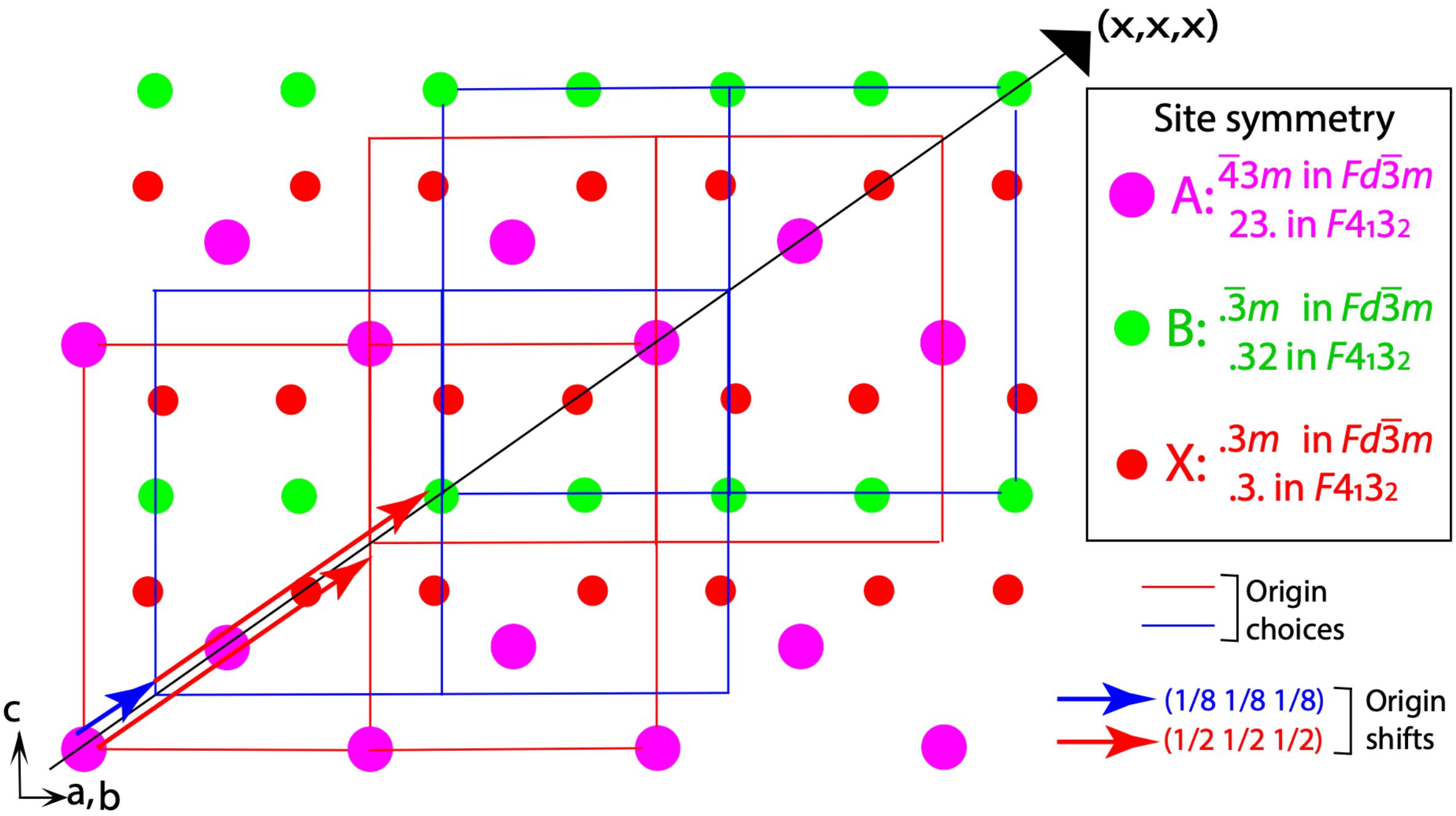

**Figure 3.** Representative cross-section of the cubic $AB_2X_4$ spinel structure. Four possible origin choices of the same conventional structure presentation are shown by four projections of the corresponding unit cells along the [1 -1 0] direction. The A, B, and X independent atoms are located on the indicated 3-fold axis. Their local symmetry (shown) is independent of the chosen structure representation options; it is systematically lower in $F4_132$ compared to $Fd\bar{3}m$. Origin shifts between the specified cells are indicated by arrows. Additional characteristics of the choice of coordinate origin and corresponding sets of atoms are given in Table 6.

**Table 1** Comparison of the $Fd\bar{3}m$ and $F4_132$ space groups describing the structure of spinel $AB_2X_4$.

Characteristics of space groups are taken from the International Tables for Crystallography (2002), Volume A "Space Group Symmetry", Fifth edition).

| **Space group** | **$Fd\bar{3}m$ (no. 227); origin choice 1 at $\bar{4}3m$** | **$F4_132$ (no.210); origin choice at 23.** |
|---|---|---|
| General reflection conditions | $hkl$*:<br>$h+k=2n$, $h+l=2n$, $k+l=2n$;<br>$0kl$*:<br>**$k+l=4n$**, $k,l=2n$;<br>$hhl$*:<br>$h+l=2n$;<br>$h00$*:<br>$h=4n$ | $hkl$*:<br>$h+k=2n$, $h+l=2n$, $k+l=2n$;<br>$0kl$*:<br>$k,l=2n$;<br>$hhl$*:<br>$h+l=2n$;<br>$h00$*:<br>$h=4n$ |

| **Characteristics of A site, 8*a*. Lattice complex: 8*a* $\bar{4}3m$ $Fd\bar{3}m$ *a D*** | | |
|---|---|---|
| **Space group** | **$Fd\bar{3}m$ (no. 227); origin choice 1 at $\bar{4}3m$** | **$F4_132$ (no.210); origin choice at 23.** |
| *x,y,z* (fixed) | 000, (¾ ¼ ¾)* | 000, (¾ ¼ ¾)* |
| Site symmetry | $\bar{4}3m$ | 23. |
| Extra reflection conditions | *hkl**: <br> *h* =2*n*+1, or *h*+*k*+*l* = 4*n* | *hkl**: <br> *h* =2*n*+1, or *h*+*k*+*l* = 4*n* (*0kl**: ***k+l = 4n)*** |
| **Characteristics of B site, 16*d*. Lattice complex: 16*d* $.\bar{3}m$ $Fd\bar{3}m$ *c T*** | | |
| **Space group** | **$Fd\bar{3}m$ (no. 227); origin choice 1 at $\bar{4}3m$** | **$F4_132$ (no.210); origin choice at 23.** |
| *x,y,z* (fixed) | 5/8 5/8 5/8, (3/8 7/8 1/8)* | 5/8 5/8 5/8, (3/8 7/8 1/8)* |
| Site symmetry | $.\bar{3}m$ | .32 |
| Extra reflection conditions | *hkl**: <br> *h* =2*n*+1, or *h,k,l* =4*n*+2, or *h,k,l* = 4*n* | *hkl**: <br> *h* =2*n*+1, or *h,k,l* = 4*n*+2, or *h,k,l* = 4*n* (*0kl**: ***k+l* = 4*n*)** |
| **Characteristics of X site, 32*e*. Lattice complex: 32*e* .3*m* $Fd\bar{3}m$ *e* ..2*D*4*xxx*** | | |
| **Space group** | **$Fd\bar{3}m$ (no. 227); origin choice 1 at $\bar{4}3m$** | **$F4_132$ (no.210); origin choice at 23.** |
| *x,x,x* (*x* is variable) | *xxx*, (-*x* -*x*+½ *x*+½)* <br> -*x*+¼ -*x*+¼ -*x*+¼, (*x*+¼ -*x*+¾ *x*+¾)* | *xxx*, (-*x* -*x*+½ *x*+½)* <br> -*x*+¼ -*x*+¼ -*x*+¼, (*x*+¼ -*x*+¾ *x*+¾)* |
| Site symmetry | .3*m* | .3. |
| Extra reflection conditions | No extra conditions | *0kl**: *k+l = 4n* |
| Reflection conditions for $AB_2O_4$ spinel structure type | *hkl**: <br> *h*+*k* =2*n*, *h*+*l* =2*n*, *k*+*l* = 2*n*; <br> *0kl**: <br> *k*+*l* = 4*n*, *k,l* = 2*n*; <br> *hhl**: <br> *h*+*l* = 2*n*; <br> *h00**: <br> *h* = 4*n* | *hkl**: <br> *h*+*k* =2*n*, *h*+*l* =2*n*, *k*+*l* = 2*n*; <br> *0kl**: <br> *k*+*l* = 4*n*, *k,l* = 2*n*; <br> *hhl**: <br> *h*+*l* = 2*n*; <br> *h00**: <br> *h* = 4*n* |

* The cyclically permuted coordinates *x,y,z* and indexes *h,k,l* are also included.

**Table 2** Comparison of ADP parameters up to the 3rd order of the tensor in the $Fd\bar{3}m$ and $F4_132$ space groups describing the structure of spinel $AB_2X_4$. $U_{jl}$ and $C_{jkl}$ define the parameters of the harmonic tensor and the 3rd order ADP tensor, respectively. Independent parameters are indicated in bold.

| **Atom** | $Fd\bar{3}m$ | $F4_132$ |
|---|---|---|
| A | $\mathbf{U_{11}} = U_{22} = U_{33} \neq 0$<br>$\mathbf{U_{12}} = U_{13} = U_{23} = 0$ | $\mathbf{U_{11}} = U_{22} = U_{33} \neq 0$<br>$\mathbf{U_{12}} = U_{13} = U_{23} = 0$ |
| | $\mathbf{C_{111}} = C_{112} = C_{113} = C_{122} = C_{222} = C_{133} = C_{223} = = C_{233} = C_{333} = 0$<br>$\mathbf{C_{123}} \neq 0$ | $\mathbf{C_{111}} = C_{112} = C_{113} = C_{122} = C_{222} = C_{133} = C_{223} = = C_{233} = C_{333} = 0$<br>$\mathbf{C_{123}} \neq 0$ |
| B | $\mathbf{U_{11}} = U_{22} = U_{33} \neq 0$<br>$\mathbf{U_{12}} = U_{13} = U_{23} \neq 0$ | $\mathbf{U_{11}} = U_{22} = U_{33} \neq 0$<br>$\mathbf{U_{12}} = U_{13} = U_{23} \neq 0$ |
| | $\mathbf{C_{111}} = C_{222} = C_{333} = C_{123} = C_{113} = C_{112} = C_{122} = = C_{133} = C_{223} = C_{233} = 0$ | $\mathbf{C_{111}} = C_{222} = C_{333} = C_{123} = 0$<br>$\mathbf{C_{113}} = C_{122} = C_{233} = -C_{112} = -C_{133} = -C_{223} \neq 0$ |
| X | $\mathbf{U_{11}} = U_{22} = U_{33} \neq 0$<br>$\mathbf{U_{12}} = U_{13} = U_{23} \neq 0$ | $\mathbf{U_{11}} = U_{22} = U_{33} \neq 0$<br>$\mathbf{U_{12}} = U_{13} = U_{23} \neq 0$ |
| | $\mathbf{C_{111}} = C_{222} = C_{333} \neq 0$<br>$\mathbf{C_{113}} = C_{112} = C_{122} = C_{133} = C_{233} = C_{223} \neq 0$<br>$\mathbf{C_{123}} \neq 0$ | $\mathbf{C_{111}} = C_{222} = C_{333} \neq 0$<br>$\mathbf{C_{113}} = C_{233} = C_{122} \neq 0$<br>$\mathbf{C_{112}} = C_{133} = C_{223} \neq 0$<br>$\mathbf{C_{123}} \neq 0$ |

**Table 3** Details of the $LiMn_2O_4$ crystal structure simulation in two space groups, $Fd\bar{3}m$ and $F4_132$, using Co $K\alpha$ radiation ($\lambda$ = 1.79027 Å) and Mo $K\alpha$ radiation ($\lambda$ = 0.70926 Å).

Using the Co radiation: the anomalous dispersion components for Li – $f'$ = 0.001, $f''$ = 0.001, for Mn – $f'$ = -2.079, $f''$ = 3.555 and for O – $f'$ = 0.063, $f''$ = 0.044. Using the Mo radiation: the anomalous dispersion components for Li – $f'$ = 0.000, $f''$ = 0.000, for Mn – $f'$ = 0.337, $f''$ = 0.728 and for O – $f'$ = 0.011, $f''$ = 0.006.

| | **Co *K*a radiation** | | **Mo *K*a radiation** | |
|---|---|---|---|---|
| Chemical formula | $LiMn_2O_4$ | $LiMn_2O_4$ | $LiMn_2O_4$ | $LiMn_2O_4$ |
| Crystal system, space group | Cubic, $Fd\bar{3}m$ (no. 227) | Cubic, $F4_132$ (no. 210) | Cubic, $Fd\bar{3}m$ (no. 227) | Cubic, $F4_132$ (no. 210) |
| Origin choice | $\bar{4}3m$ (Choice 1) | .32 | $\bar{4}3m$ (Choice 1) | .32 |
| Shift of origins | 000 | 000 | 000 | 000 |

| Temperature (K) | 293 | 293 | 293 | 293 |
|---|---|---|---|---|
| *a* (Å) | 8.2261 (2) | 8.2261 (2) | 8.2261 (2) | 8.2261 (2) |
| *V* ($Å^3$) | 556.65 (2) | 556.65 (2) | 556.65 (2) | 556.65 (2) |
| *Z* | 8 | 8 | 8 | 8 |
| No. of reflections with $I > 3\sigma(I)$ | 760 | 760 | 1428 | 1428 |
| $(\sin \theta / \lambda)_{max}$ ($Å^{-1}$) | 0.558 | 0.558 | 0.700 | 0.700 |
| Range of *h*, *k*, *l* | $h$ = -9→9,<br>$k$ = -9→9,<br>$l$ = -9→9 | $h$ = -9→9,<br>$k$ = -9→9,<br>$l$ = -9→9 | $h$ = -11→11,<br>$k$ = -11→11,<br>$l$ = -11→11 | $h$ = -11→11,<br>$k$ = -11→11,<br>$l$ = -11→11 |
| Atom (Wyckoff site): coordinates | Li (8*a*): 000<br>Mn (16*d*):<br>5/8 5/8 5/8<br>O (32e): 0.3888 0.3888 0.3888 | Li (8*a*): 000<br>Mn (16*d*):<br>5/8 5/8 5/8<br>O (32e): 0.3888 0.3888 0.3888 | Li (8*a*): 000<br>Mn (16*d*):<br>5/8 5/8 5/8<br>O (32e): 0.3888 0.3888 0.3888 | Li (8*a*): 000<br>Mn (16*d*):<br>5/8 5/8 5/8<br>O (32e): 0.3888 0.3888 0.3888 |
| ADP | Li:<br>$U_{11}$ = 0.021807,<br>$C_{123}$ = 0.001834<br>Mn:<br>$U_{11}$ = 0.010834,<br>$U_{12}$ = 0.00151,<br>**$C_{113}$ = 0**<br>O:<br>$U_{11}$ = 0.015756,<br>$U_{12}$ = 0.00224,<br>$C_{111}$ = 0.002712,<br>**$C_{112}$ = 0**,<br>$C_{113}$ = -0.000956,<br>$C_{123}$ = -0.000231 | Li:<br>$U_{11}$ = 0.021807,<br>$C_{123}$ = 0.001834<br>Mn:<br>$U_{11}$ = 0.010834,<br>$U_{12}$ = 0.00151,<br>**$C_{113}$ = -0.001088**<br>O:<br>$U_{11}$ = 0.015756,<br>$U_{12}$ = 0.00224,<br>$C_{111}$ = 0.002712,<br>**$C_{112}$ = -0.001799**<br>$C_{113}$ = -0.000109,<br>$C_{123}$ = 0.00089 | Li:<br>$U_{11}$ = 0.021807,<br>$C_{123}$ = 0.001834<br>Mn:<br>$U_{11}$ = 0.010834,<br>$U_{12}$ = 0.00151,<br>**$C_{113}$ = 0**<br>O:<br>$U_{11}$ = 0.015756,<br>$U_{12}$ = 0.00224,<br>$C_{111}$ = 0.002712,<br>**$C_{112}$ = 0**,<br>**$C_{113}$** = -0.000956,<br>$C_{123}$ = -0.000231 | Li:<br>$U_{11}$ = 0.021807,<br>$C_{123}$ = 0.001834;<br>Mn:<br>$U_{11}$ = 0.010834,<br>$U_{12}$ = 0.00151,<br>**$C_{113}$ = -0.001088**<br>O:<br>$U_{11}$ = 0.015756,<br>$U_{12}$ = 0.00224,<br>$C_{111}$ = 0.002712,<br>**$C_{112}$ = -0.001799**<br>$C_{113}$ = -0.000109,<br>$C_{123}$ = 0.00089 |

Only independent ADP parameters are shown. The parameters that differ fundamentally in space groups $Fd\bar{3}m$ and $F4_132$ are indicated in bold.

**Table 4** Comparison of the squared structure factors amplitudes, $|F(\mathbf{h})|^2$ and $|F(\mathbf{-h})|^2$, for a few representative reflections of the $LiMn_2O_4$ structure simulated in space groups $Fd\bar{3}m$ and $F4_132$ using Mo and Co radiations and the 3rd order ADP tensor.

| **(*h k l*)** | **Co-radiation** | | **Mo-radiation** | |
|---|---|---|---|---|
| | $\|F(\mathbf{h})\|^2$ and $\|F(\mathbf{-h})\|^2$ | $\Delta\|F\|^2$, % | $\|F(\mathbf{h})\|^2$ and $\|F(\mathbf{-h})\|^2$ | $\Delta\|F\|^2$, % |
| | $Fd\bar{3}m$ | | $Fd\bar{3}m$ | |
| (1 5 7) & (-1 -5 -7) | 3081.2 & 3081 | 0 | 4364.6 & 4364.6 | 0 |
| (1 3 5) & (-1 -3 -5) | 8630.7 & 8630.7 | 0 | 11331.8 & 11331.8 | 0 |
| (2 4 6) & (-2 -4 -6) | 193.9 & 193.9 | 0 | 199.3 & 199.3 | 0 |
| (2 4 8) & (-2 -4 -8) | 82.1 & 82.1 | 0 | 62.3 & 62.3 | 0 |
| | $F4_132$ | | $F4_132$ | |
| (1 5 7) & (-1-5-7) | 3105.6 and 3112.9 | 0.2 | 3242.6 and 3243.7 | 0.03 |
| (1 3 5) & (-1 -3 -5) | 8678.7 and 8686.7 | 0.09 | 8396.1 and 8397.2 | 0.01 |
| (2 4 6) & (-2 -4 -6) | 222.8 and 180.0 | 21.2 | 156.6 and 150.2 | 4.2 |
| (2 4 8) & (-2 -4 -8) | 99.7 and 72.6 | 10.7 | 51.6 and 47.6 | 8.1 |

**Table 5** Atomic coordinates of $Li(Mn,Ni)_2O_4$ in $Fd\bar{3}m$ and $F4_132$ space groups after Amin *et al.* (2020).

| Crystal system, space group | Cubic, $Fd\bar{3}m$ (no. 227) | Cubic, $F4_132$ (no. 210) |
|---|---|---|
| Origin choice | Not indicated | Not indicated |
| Atom (Wyckoff site): coordinates | Li (8*b*): 3/8 3/8 3/8<br>(Mn,Ni) (16*c*): 000<br>O (32*e*): 0.2368 0.2368 0.2368 | Li (8*a*): ½ ½ ½<br>(Mn,Ni) (16*d*): 1/8 1/8 1/8<br>O (32*e*): 0.3618 0.3618 0.3618 |

**Table 6** Atomic positions in four possible unit cells characteristic of the same $AB_2X_4$ cubic spinel structure in both $Fd\bar{3}m$ and $F4_132$ space groups.

For better comparison with Figure 3, only coordinates of type (*xxx*) located on one of the four axes of threefold symmetry are indicated. Coordinates listed in International Tables for Crystallography are underlined in bold.

| | **Origin choice 1 at $\bar{4}3m$ in $Fd\bar{3}m$, accordingly at 23. in $F4_132$** | | **Origin choice 2 at $.\bar{3}m$ in $Fd\bar{3}m$, accordingly at .32 in $F4_132$** | |
|---|---|---|---|---|
| Origin shift | (000) | (½ ½ ½) | (1/8 1/8 1/8) | (1/8 1/8 1/8) + |

| | | | | (½ ½ ½) = (5/8 5/8 5/8) |
|---|---|---|---|---|
| Coordinate shift | (000) | (-½ -½ -½) | (-1/8 -1/8 -1/8) | (-5/8 -5/8 -5/8) |
| A | **8*a*: (0,0,0)**; (¼,¼,¼) | **8*b*: (½,½,½)**; (¾,¾,¾) | **8*a***: (7/8,7/8,7/8); **(1/8,1/8,1/8)** | **8*b*: (3/8,3/8,3/8);** (5/8,5/8,5/8) |
| B | **16*d*: (5/8,5/8,5/8)** | **16*c*: (1/8,1/8,1/8)** | **16*d*: (½,½,½ )** | **16*c*: (0,0,0)** |
| X | **32*e*:** **($x_1$,$x_1$,$x_1$);** (5/4-$x$1,5/4-$x$1,5/4-$x$1) = (¼ -$x_1$,¼-$x_1$,¼-$x_1$) | **32*e*:** ($x_1$-½,$x_1$-½,$x_1$-½); (¾-$x$1,¾-$x$1,¾-$x$1) | **32*e*:** ($x_1$-1/8,$x_1$-1/8,$x_1$-1/8); (1/8-$x_1$,1/8-$x_1$,1/8-$x_1$) | **32*e*:** ($x_1$-5/8,$x_1$-5/8,$x_1$-5/8); (5/8-$x_1$,5/8-$x_1$,5/8-$x_1$) |
| X = O in $Li(Mn,Ni)_2O_4$ | **32*e*:** (0.862,0.862,0.862); (0.388,0.388,0.388) | **32*e*:** (0.362,0.362,0.362); (0.888,0.888,0.888) | **32*e*:** (0.737,0.737,0.737); (0.263,0.263,0.263) | **32*e*:** (0.237,0.237,0.237); (0.763,0.763,0.763) |

**Acknowledgements** This work was supported by the Swiss National Science Foundation (SNSF) Sinergia network NanoSkyrmionics (grant No. CRSII5-171003).

**Conflicts of interest** There are no conflicts of interest.